\documentclass{article}
\usepackage{spconf,amsmath,graphicx}
\usepackage{cite}
\usepackage{amssymb}
\usepackage{multirow}
\usepackage{booktabs}
\usepackage{color}
\usepackage{caption}

\title{Robust data2vec: Noise-robust speech representation learning for ASR by combining regression and improved contrastive learning}
\name{Qiu-Shi Zhu$^1$, Long Zhou$^2$, Jie Zhang$^{1}$, Shu-Jie Liu$^2$, Yu-Chen Hu$^3$, Li-Rong Dai$^1$
	\thanks{This work is supported by the National Natural Science Foundation of China (62101523), Hefei Municipal Natural Science Foundation (2022012), 
		Fundamental Research Funds for the Central Universities and the Leading Plan of CAS (XDC08010200).}}
\address{ $^1$NERC-SLIP, University of Science and Technology of China (USTC), Hefei, China\\
	$^2$Microsoft Research Asia\\ 
	$^3$Nanyang Technological University, Singapore
}

\begin{document}
	\ninept
	\maketitle
	\begin{abstract}
		Self-supervised pre-training methods based on contrastive learning or regression tasks can utilize more unlabeled data to improve the performance of automatic speech recognition (ASR).
		However, the robustness impact of combining the two pre-training tasks and constructing different negative samples for contrastive learning still remains unclear.
        In this paper, we propose a noise-robust data2vec for self-supervised speech representation learning by jointly optimizing the contrastive learning and regression tasks in the pre-training stage.
        Furthermore, we present two improved methods to facilitate contrastive learning.
		More specifically, we first propose to construct patch-based non-semantic negative samples to boost the noise robustness of the pre-training model, which is achieved by dividing the features into patches at different sizes (i.e., so-called negative samples).
		Second, by analyzing the distribution of positive and negative samples, we propose to remove the easily distinguishable negative samples to improve the discriminative capacity for pre-training models.
		Experimental results on the CHiME-4 dataset show that our method is able to improve the performance of the pre-trained model in noisy scenarios.
		We find that joint training of the contrastive learning and regression tasks can avoid the model collapse to some extent compared to only training the regression task.

	\end{abstract}
	\section{Introduction}
	\label{sec:intro}
	Collecting labeled data is rather time-consuming and economically expensive, while there exists a large amount of unlabeled data that can be recorded in reality. How to make a better advantage of unlabeled data for supervised learning has thus become a hot spot recently.
	In the speech community, many methods have been proposed to improve the automatic speech recognition (ASR) performance using unlabeled speech data, such as self-supervised pre-training and teacher-student schemes (i.e., so-called self-training), which were shown to be beneficial for various downstream speech tasks \cite{NEURIPS2020_92d1e1eb,9585401,9814838,ao2021speecht5,baevski2022data2vec,zhang2022speechlm}.
	For example, the wav2vec2.0~\cite{NEURIPS2020_92d1e1eb}  based on self-supervised pre-training utilizes a contrastive loss function to narrow the distance between the predicted and positive samples and meanwhile enlarge the distance between the predicted and negative samples.
	In wav2vec2.0~\cite{NEURIPS2020_92d1e1eb}, local features are employed as targets for self-supervised pre-training, where the contextual information is not fully leveraged. This problem was then considered in HuBERT~\cite{9585401}, which offline clusters the representations output from the middle layer of the pre-trained model to generate targets for self-supervised pre-training. On the basis of HuBERT,
	WavLM~\cite{9814838}  utilizes a sentence-level mixing data augmentation approach to enhance the speaker information, which performs very well on the SUPERB benchmark~\cite{yang21c_interspeech}.
	Unlike wav2vec2.0 and HuBERT which employ local information and discrete contextual information as targets, data2vec~\cite{baevski2022data2vec} follows the teacher-student scheme~\cite{NIPS2017_68053af2,zhang2022cross} and adopts continuous contextual representations as targets to preform regression tasks, leading to an even better performance on downstream tasks.

	It was  shown that self-supervised pre-training can improve the noise robustness of ASR models. For example, problem-agnostic speech encoder (PASE+)~\cite{ravanelli2020multi} uses an online speech perturbation module and employs multiple self-supervised tasks to improve the noise robustness.
	In robust wav2vec2.0~\cite{hsu2021robust},  a more general case is explored, where the domain of unlabeled data used for pre-training is different from that of the labeled data for fine-tuning, exhibiting a stronger generalization capacity for ASR models.
	By using quantized clean speech features as pre-training targets, enhanced wav2vec2.0~\cite{zhu2022noise} can improve the noise robustness of ASR models.
	Wav2vec-switch~\cite{wang2021wav2vec} allows the model to have consistent predictions for both original and noisy speech by contrastive learning.
	In~\cite{wang2021improving}, a reconstruction module was proposed based on wav2vec2.0 to improve the noise robustness of the learned representations.
    However, there is few work to investigate the speech representation robustness of combining regression and contrastive learning tasks, especially the different negative samples in contrastive learning. 
	
	In the field of computer vision (CV), many analytical works have been done on negative samples in contrastive learning, and it was shown that negative samples affect the quality of the pre-trained representations.
	In~\cite{NEURIPS2020_f7cade80}, it was found that many negative samples are too far away from positive samples within contrastive learning, and hard negative mixing~\cite{NEURIPS2020_f7cade80} was therefore proposed, which can improve the performance and training efficiency of pre-trained models by mixing difficult negative samples at the feature level.
	To alleviate the sampling bias of negative samples in contrastive learning, debiased contrastive loss was proposed in~\cite{NEURIPS2020_63c3ddcc}.
	In~\cite{robinson2021contrastive}, based on the fact that the representation of contrastive learning benefits from hard negative samples, a hard negative samples selection approach was proposed, in which the user can control the difficulty of the negative samples.
	In~\cite{cai2020all}, it was shown that  only 5\% of the hardest negative samples are both useful and sufficient for downstream tasks,  95\% of the negative samples are unnecessary, and 0.1\% of the hardest are even harmful.
	In addition, a false negative detection method was proposed in~\cite{chen2022incremental} to address the problem that positive samples may be erroneously treated as negative samples.
	To further improve the robustness of pre-trained models, texture- and patch-based negative sample construction methods were proposed in~\cite{NEURIPS2021_e5afb0f2}.
 Speech processing tasks, such as ASR, require negative samples at the frame level, which is different from typical classification tasks in CV,
	it is thus necessary to analyze the impact of negative samples in contrastive learning for speech-related downstream tasks.

    In this paper, we propose a non-semantic negative sample construction based noise-robust speech representation learning model, namely robust data2vec, which is jointly pre-trained with the regression and contrastive learning tasks.
	Inspired by~\cite{NEURIPS2021_e5afb0f2}, we first propose a patch-based non-semantic negative samples construction method that enables the model to learn high-level noise-robust representations. 
	By analyzing the distribution of positive and negative samples of the standard data2vec~\cite{baevski2022data2vec} model, we find that there are a large number of negative samples that are easily discriminated by the model in the late stage of model training, which means that these samples do not play a role.
	Hence, we perform an incremental removal of the easier negative samples by using the cosine similarity scores between positive and negative samples.
  Experimental results show that the proposed method can improve the noise robustness of the model in noisy scenarios.
	We conclude that jointly optimizing the regression and contrastive learning tasks can reduce the possibility of representation collapse in the teacher-student framework compared to the regression-only task.
	Our pre-trained models and code are available at https://github.com/zqs01/data2vecnoisy.

	\begin{figure}[!t]
		\centering
		\includegraphics[width=0.4\textwidth]{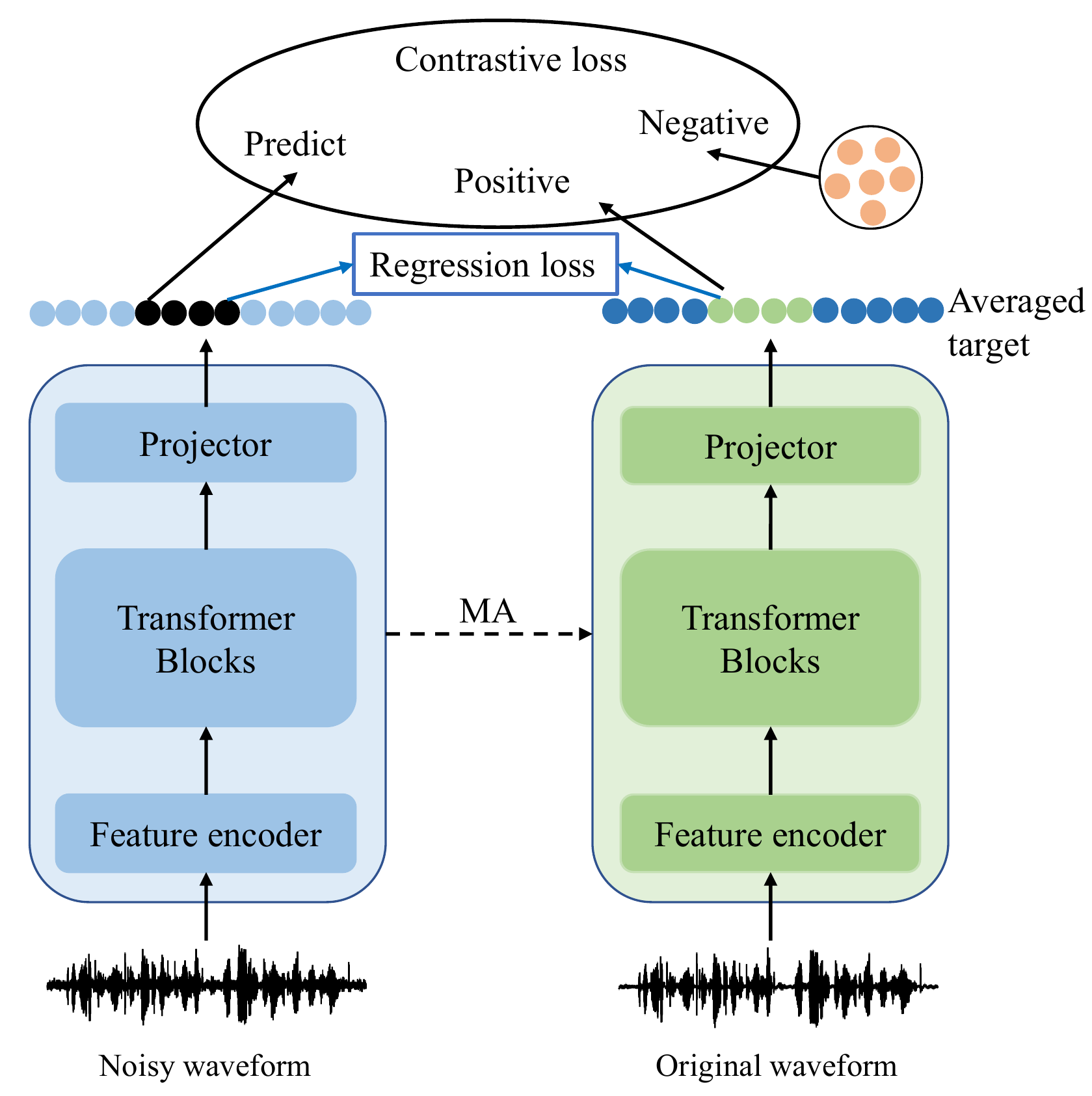}
		\caption{The structure of robust data2vec for speech representation learning with regression loss and improved contrastive loss.}
		\vspace{-0.4cm}
		\label{fig:figure2}
	\end{figure}

 \vspace{-0.15cm}
\section{Methodology}
\vspace{-0.15cm}

In this section, we first introduce the overall framework of robust data2vec, which is jointly trained with a regression task and noise-robust contrastive learning task (Section \ref{sec:framework}). Then we present the two improved methods for constrastive learning, including constracting non-semantic negative samples (Section \ref{sec:constructing}), and removing distinguishable negative samples (Section \ref{sec:removing}).
	\vspace{-0.15cm}
	\label{sec:method}

	\subsection{Joint training regression and contrastive learning tasks}
        \label{sec:framework}
	The overall model structure we utilize is shown in Fig.~\ref{fig:figure2}, which is similar to the data2vec~\cite{baevski2022data2vec}.
	The model contains a feature encoder and a multi-layer Transformer encoder, where the feature encoder consists of seven convolution layers.
	The left side depicts the student model, and  the teacher model at the right side is obtained by moving average (MA) of the student model and provides targets for model training.
	Letting $\theta$ denote the training parameter of the student model, the teacher model parameter $\Delta$ can be obtained by
	$\Delta \leftarrow \tau\Delta + (1-\tau)\theta$, where $\tau$ denotes a linear increase from $\tau_{0}$ to $\tau_{e}$ at an increment of $\tau_{n}$ and remains constant thereafter.
	This strategy allows the model to be updated faster in early stages and slower in later stages when training.
	Specifically, the input of the student model is the noisy speech waveform $x_{\rm noisy}$, and the input of the teacher model is the original speech waveform $x_{\rm origin}$, which are fed into the feature encoder to obtain the noisy feature $z_{\rm noisy}$ and the original feature $z_{\rm origin}$, respectively.
	The masked noisy and original features  are passed through the Transformer encoders at two branches to produce the predicted outputs $c_{\rm pre}$ and the target $c_{\rm origin}$, respectively.
	The output of the Transformer encoder of layer $l$ at time step $t$ is represented by $c_{t}^l$.
	We calculate the regression and contrastive losses between the predicted output $c_{\rm pre}$ and the averaged target $c_{\rm tar}$, where $c_{{\rm tar}_{t}}=1/M\sum_{l=L-M+1}^{L}c_{{\rm origin}_t}^{l}$ is the averaged output representations of the top-$M$ layer Transformer encoder.
	The regression loss $\mathcal{L}_{\rm reg}$ function adopts the same loss function as data2vec, which can be formulated as
	\begin{equation}
		\mathcal{L}_{\rm reg}=\left\{\begin{array}{ll}
			\frac{1}{2}\left(c_{{\rm pre}_{t}}-c_{{\rm tar}_{t}}\right)^{2} / \beta & \left|c_{{\rm pre}_{t}}-c_{{\rm tar}_{t}}\right| \leq \beta \\
			\left(\left|c_{{\rm pre}_{t}}-c_{{\rm tar}_{t}}\right|-\frac{1}{2} \beta\right) & \text { otherwise }
		\end{array}\right.
	\end{equation}
where $\beta$ controls the transition from squared loss to L1 loss,
	and the contrastive loss function $\mathcal{L}_c$ as
	\begin{equation}
		\mathcal{L}_{\rm c} = -\log \frac{\exp({\rm sim}(c_{{\rm pre}_t},c_{{\rm tar}_{t}})/\kappa)}{\sum_{\tilde{c} {\sim} {\{c_{\rm tar},c_n,c_{ns}\}}}\exp({\rm sim}(c_{{\rm pre}_t},\tilde{c})/\kappa)},
	\end{equation}
	resulting in the total loss function $\mathcal{L}_{\rm total}$, given by
	\begin{equation}
		\mathcal{L}_{\rm total}=\mathcal{L}_{\rm reg} + \lambda \mathcal{L}_{\rm c},
		\label{eq4} \\
	\end{equation}
	where $\lambda$ is a hyperparameter.

	\subsection{Contrastive learning with non-semantic negative samples}
        \label{sec:constructing}
	
	\begin{figure}[!t]
		\centering
		\includegraphics[width=0.45\textwidth]{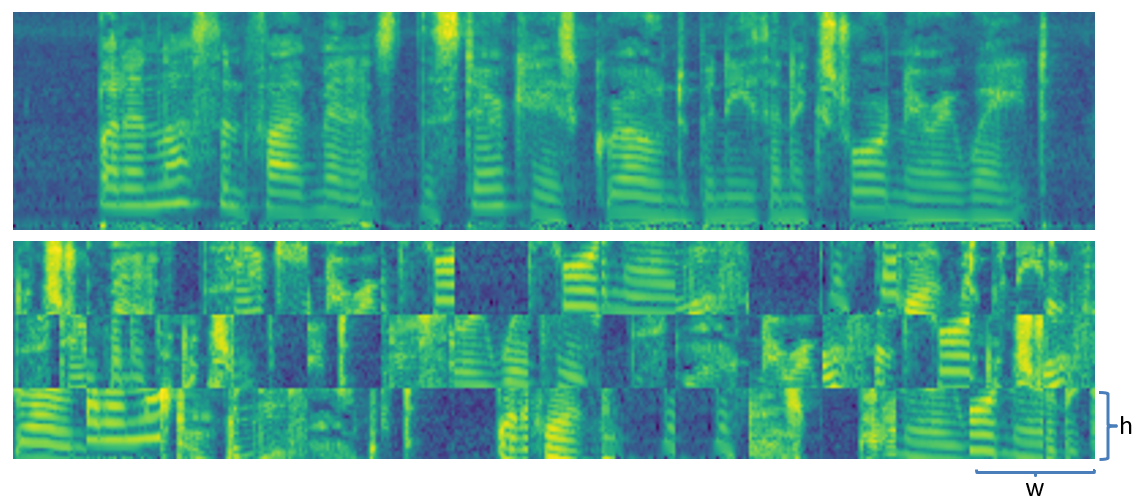}
		\caption{The top and bottom plots are  the original fbank and patch-based shuffled feature, respectively.}
		\label{fig:figure1}
		\vspace{-0.4cm}
	\end{figure}
	
	In practice, it is expected that the pre-trained model can learn high-level representations that are robust to noise-perturbed input features.
	To this end, we construct noisy and perturbed samples as negative samples (i.e., non-semantic negative samples) in the pre-training stage, such that high-level robust representations can be learned.
	In detail, we use a patch-based approach to construct non-semantic negative samples.
	The output features of the model are sliced into patches at different sizes along the time and dimensional axes and randomly shuffled.
	For instance, such patch segmentation on the fbank features is shown in Fig.~\ref{fig:figure1}, where $w$ and $h$ are hyperparameters related to the width and height of the patch, respectively. It is natural to form speech negative samples at the frame level, so that $N$ frames of standard negative samples and $N$ frames of non-semantic negative samples can be randomly selected. Standard negative samples are selected from the original features, and non-semantic negative samples are selected from the disturbed features.
	Let $c_t$, $c_p$, $c_n$ and $c_{ns}$  denote the representations of the query samples, positive samples, standard negative samples  and  non-semantic negative samples, respectively.
	The non-semantic contrastive loss function can be expressed as

	\begin{equation}
		L_{c} = -\log \frac{\exp({\rm sim}(c_{t},c_{p_{t}})/\kappa)}{\sum_{\tilde{c} {\sim} {\{c_{p},c_n,c_{ns}\}}}\exp({\rm sim}(c_{t},\tilde{c})/\kappa)},
		\label{eq1} \\
	\end{equation}
	where $\kappa$ is the temperature coefficient and $\rm sim$ denotes the cosine similarity.
	Clearly, in case the non-semantic negative samples are excluded, $L_c$ reduces to the standard contrastive loss function.

	\subsection{Removal of the more distinguishable negative samples}
        \label{sec:removing}
	
	\begin{figure}[!t]
		\centering
		\includegraphics[width=0.4\textwidth]{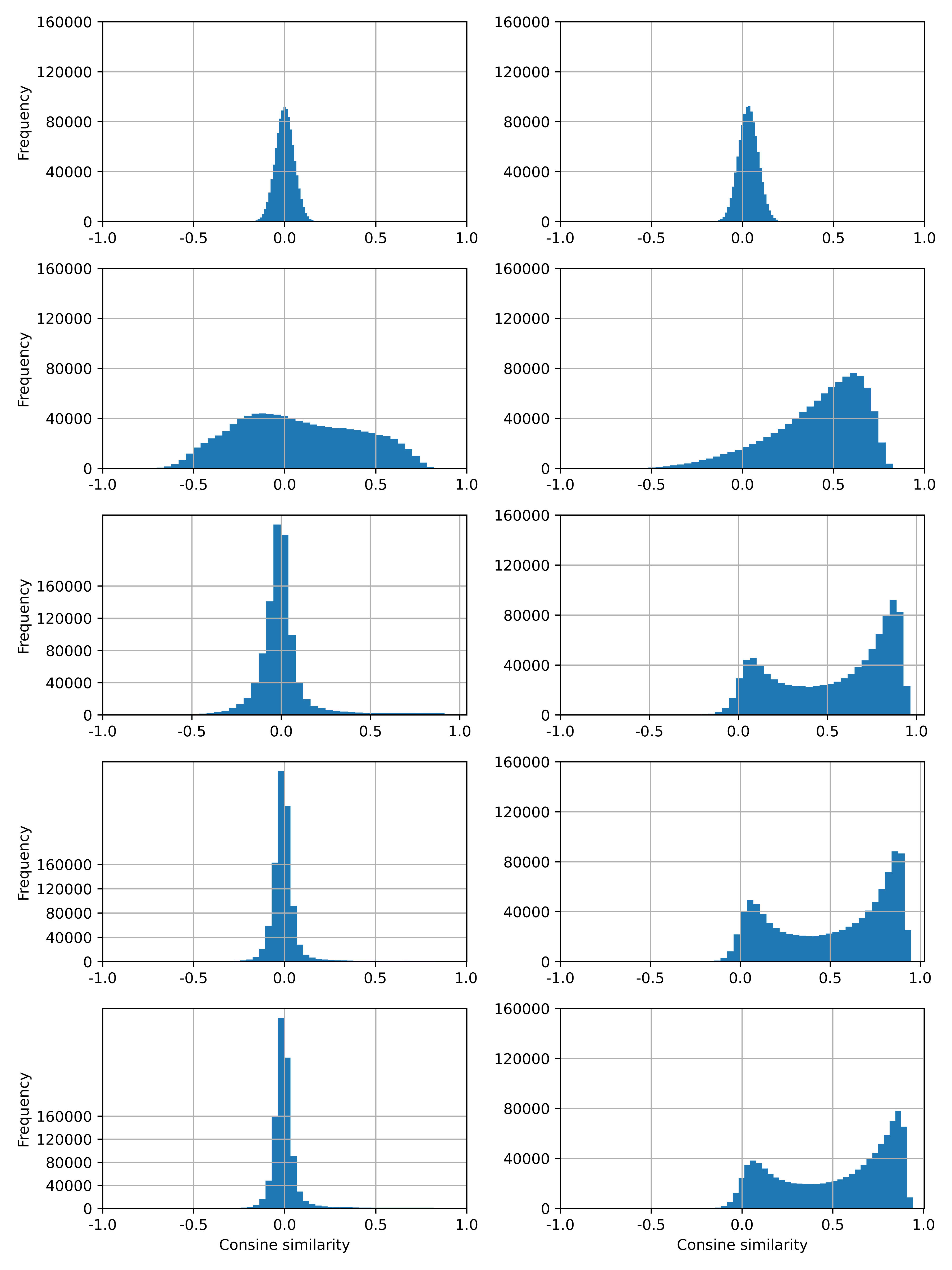}
		\caption{Cosine similarities between  1000k predicted and negative samples (left panel) and  between 1000k predicted and positive samples (right panel). From top to bottom, the plots are obtained by pre-training \{1, 100, 200, 300, 400\} epochs.}
		\vspace{-0.4cm}
		\label{fig:figure3}
	\end{figure}
	In order to remove the more distinguishable negative samples, we first analyze the distribution of positive and negative samples in contrastive learning. Taking the pre-trained data2vec~\cite{baevski2022data2vec} model as an example, we calculate the cosine similarities of the predicted-positive sample pairs and predicted-negative sample pairs obtained at different epochs, which are shown in Fig.~\ref{fig:figure3}.
	We can see  that with more pre-training epochs the model becomes more discriminative to distinguish between positive and negative samples, but the  negative samples will be distributed more sharply.
	This shows that the impact of some more distinguishable negative samples is weakened with more epochs, which can thus be removed without loss of the model capacity.
	Specifically, the negative samples with smaller predicted-negative cosine similarity scores are removed.
	In other words, we can select  $k$ out of $N$ negative samples with the highest cosine similarity  to increase the pre-training difficulty, such that more robust representations can be learned.

	\vspace{-0.15cm}
	\section{EXPERIMENTAL setup}
	\vspace{-0.15cm}

	\label{sec:experiment}
	\subsection{Data description}
	We adopt the real noisy CHiME-4~\cite{vincent20164th} dataset to validate the effectiveness of our method, which was collected by asking volunteers to read the text from  the Wall Street Journal (WSJ0) corpus using a six-channel distant microphone array and a close-by microphone.
	The dataset contains real and artificially simulated noisy speech data.
	As the focus of this work is on the single-channel ASR,  all six-channels data are used for training and the single-channel close-by noisy data are for validation as well as testing.
	The noise dataset used in the pre-training stage is derived from the MUSAN~\cite{snyder2015musan} dataset and the CHiME-4 noise dataset. The former contains 109 hours of data in total, including 60 hours of English speech data, 42 hours of music data and 6 hours of noise data.
	In order to synthesize noisy speech, the CHiME-4 data are randomly mixed with the noise data at a signal-to-noise ratio (SNR) ranging from 0 to 25 dB. Note that the noise data are not required for fine-tuning.
	\subsection{Model configuration}
	The model is implemented using the fairseq toolkit.
	The feature encoder contains seven layers of 512-dimensional convolution with kernels and strides of (10, 3, 3, 3, 3, 2, 2) and (5, 2, 2, 2, 2, 2, 2), respectively.
	The Transformer has 12 layers with 768 dimensions, where the feed-forward neural network has 3072 dimensions. 
	For joint training, we set $\tau_{0}$ = 0.999, $\tau_{e}$ = 0.9999 and $\tau_{n}$ = 30000.
	The masking strategy keeps the same as~\cite{NEURIPS2020_92d1e1eb}, where the starting time step is chosen at a probability of $p$ = 0.065 among all time steps and  the subsequent ten time-step frames are then masked.
	The model uses the Adam~\cite{kingma2014adam} optimizer at a maximum learning rate of 1e-4.
	Since the CHiME-4 data is relatively small, we use a publicly available pre-trained model\footnote{https://github.com/facebookresearch/fairseq/tree/main/examples/data2vec} for initialization and continue pre-training 100k updates at a batch size of 63 minutes of audio, which is trained using 16 Tesla-V100-32G GPUs.
	We adopt $N$ = 50 standard negative samples and $N$ = 50 non-semantic negative samples, where the negative samples are randomly picked from the output representations of the teacher model.
	Both $w$ and $h$ are randomly distributed in [30, 50].
	The outputs of the top eight (i.e., $M$ = 8) layers of the model are averaged  as targets.
	Besides, in (\ref{eq4}) $\lambda$ is set to be 1.0.
	
	After pre-training, a linear layer is added on the top of the Transformer encoder, and the student model can then be fine-tuned using the labeled CHiME-4 data.
	The model adopts the Connectionist Temporal Classification (CTC)~\cite{kim2017joint} loss function, where the modeling units contain 26 letters and 4 special symbols.
	The performance on the test and validation sets is evaluated in terms of the word error rate (WER).
	More details can be found from our code.
	\vspace{-0.15cm}
	\section{Experimental Results}
	\vspace{-0.15cm}
	\label{sec:result}
	
	\begin{table}[]
		\caption{The ASR performance in WER on the CHiME-4 dataset.}
		\label{tab:table1}
		\centering
		\begin{tabular}{lccc}
			\hline
			\multirow{2}{*}{\textbf{Model}}  & \multirow{2}{*}{\textbf{LM}} & \multicolumn{2}{c}{\textbf{WER}}          \\ \cline{3-4} 
			&                              & \textbf{dt05\_real} & \textbf{et05\_real} \\ \hline \hline
			\multicolumn{4}{l}{\textbf{Supervised}}                                                                     \\
			DNN~\cite{menne2016rwth} & N-gram & 11.6 & 23.7   \\
			Du et al.~\cite{du2016ustc}                        & LSTM                         & 4.5                 & 9.2                 \\
			Menne et al.~\cite{menne2016rwth}                     & LSTM                         & 5.1                 & 9.3                 \\
			Wang et al.~\cite{9103053}                      & LSTM                         & 3.5                 & 6.8                 \\ \hline \hline
			\textbf{Self-supervised}         &                              &                     &                     \\ 
			Wang et al.(Large)~\cite{wang2021improving}               & LSTM                         & 2.8                 & 5.8                 \\ \hline
			Wav2vec2.0 Base~\cite{gao2021data} & None & 10.3 & 17.8 \\
			Gao et al.~\cite{gao2021data} & None  & 8.7 & 15.8 \\ \hline
			\multirow{2}{*}{Wav2vec2.0 Base~\cite{wang2021wav2vec}} & None                         & 10.6                & 17.6                \\
			& LSTM                         & 3.7                 & 7.2                 \\
			\multirow{2}{*}{Wav2vec-switch~\cite{wang2021wav2vec}}  & None                         & 10.0                & 16.5                \\
			& LSTM                         & 3.5                 & 6.6                 \\ \hline
			\multirow{2}{*}{HuBERT Base~\cite{zhu2022joint}}     & None                         & 10.4                & 17.0                \\
			& LSTM                         & 3.8                 & 7.1                 \\ 
			\multirow{2}{*}{Wav2vec2.0 Base~\cite{zhu2022joint}} & None                         & 10.5                & 17.3                \\
			& LSTM                         & 3.8                 & 7.5                 \\ 
			\multirow{2}{*}{\begin{tabular}[c]{@{}l@{}}Enhanced\\ wav2vec2.0~\cite{zhu2022joint}\end{tabular}} & None                         & 9.4                & 15.6                \\
			& LSTM                         & 3.5                 & 6.4                 \\ \hline
			\multirow{2}{*}{Data2vec Base}   & None                         & 9.5                & 15.7                \\
			& Transformer                  & 3.5                    &  6.5                   \\ \hline
			\multirow{2}{*}{Robust data2vec (\textbf{Ours})}            & None                         & 8.3                 & 12.8                \\
			& Transformer                  & \textbf{3.1}                    &  \textbf{5.8}                   \\ \hline
		\end{tabular}
	\vspace{-0.3cm}
	\end{table}
	
	\textbf{Comparison methods} In this section,  both supervised and self-supervised methods on the CHiME-4 dataset will be compared for completeness. 
	For the supervised method, the ASR model~\cite{9103053} based on speech enhancement can achieve WERs of 3.5 and 6.8 on the  validation and test sets, respectively.
	For self-supervised pre-training methods, wav2vec2.0 is based on contrastive learning, where local features are quantized and are used as targets for pre-training.
	HuBERT is based on masking-prediction, where intermediate layer representations are clustered to provide targets for pre-training.
	Data2vec is based on the teacher-student framework, where the targets for pre-training are the continuous context representations.
	Some other variants of these models are also compared.

 \textbf{Main results} The results on the CHiME-4 dataset are shown in Table~\ref{tab:table1}.
	Without a language model (LM), the wav2vec2.0 model in~\cite{wang2021wav2vec} obtains a WER of 10.6/17.6 on the validation/test set, while the WER of wav2vec-switch~\cite{wang2021wav2vec} is 10.0/16.5.
	The pre-trained wav2vec2.0~\cite{zhu2022joint} achieves a WER of 10.5/17.3, which is comparable to~\cite{wang2021wav2vec}.
	Compared with wav2vec2.0, the HuBERT and data2vec models perform slightly better in noisy scenarios.
	It is clear that the proposed method can further decrease the WER to  8.3/12.8 at the absence of LM.

	\begin{table}[]
		\caption{Results with different pre-training loss functions.}
		\label{tab:table2}
		\centering
		\begin{tabular}{l|cc}
			\hline
			\multirow{2}{*}{\textbf{Configuration}}             & \multicolumn{2}{c}{\textbf{WER}}          \\ \cline{2-3} 
			& \textbf{dt05\_real} & \textbf{et05\_real} \\ \hline
			$L_{\rm reg}$                           & 9.5                & 15.7                \\
			$L_{\rm reg}$+$L_{\rm c}$                      & 8.9                 & 14.2                \\
			$L_{\rm reg}$+$L_{\rm c}$(non-semantic)        & 8.5                 & 13.3                \\
			$L_{\rm reg}$+$L_{\rm c}$(non-semantic)+removal & 8.3                 & 12.8                \\ \hline
		\end{tabular}
	\vspace{-0.3cm}
	\end{table}
	
	\begin{table}[]
		\caption{The WERs of different number of standard negative samples and non-semantic negative samples.}
		\label{tab:table3}
		\centering
		\begin{tabular}{ll|cc}
			\hline
			\multirow{2}{*}{\begin{tabular}[c]{@{}l@{}}\textbf{Standard}\\ \textbf{negatives}\end{tabular}} & \multirow{2}{*}{\begin{tabular}[c]{@{}l@{}}\textbf{Non-semantic}\\ \textbf{negatives}\end{tabular}} & \multicolumn{2}{c}{\textbf{WER}} \\ \cline{3-4} 
			&                                                                                   & \textbf{dt05\_real} & \textbf{et05\_real} \\ \hline
			N=100                                                                         & N=0                                                                               & 8.9        & 14.2       \\
			N=80                                                                          & N=20                                                                              & 8.6        & 13.6       \\
			N=50                                                                          & N=50                                                                              & 8.5        & 13.3       \\
			N=20                                                                          & N=80                                                                              & 8.7        & 13.9       \\ \hline
		\end{tabular}
	\end{table}

	\begin{figure}[!t]
		\centering
		\includegraphics[width=0.45\textwidth]{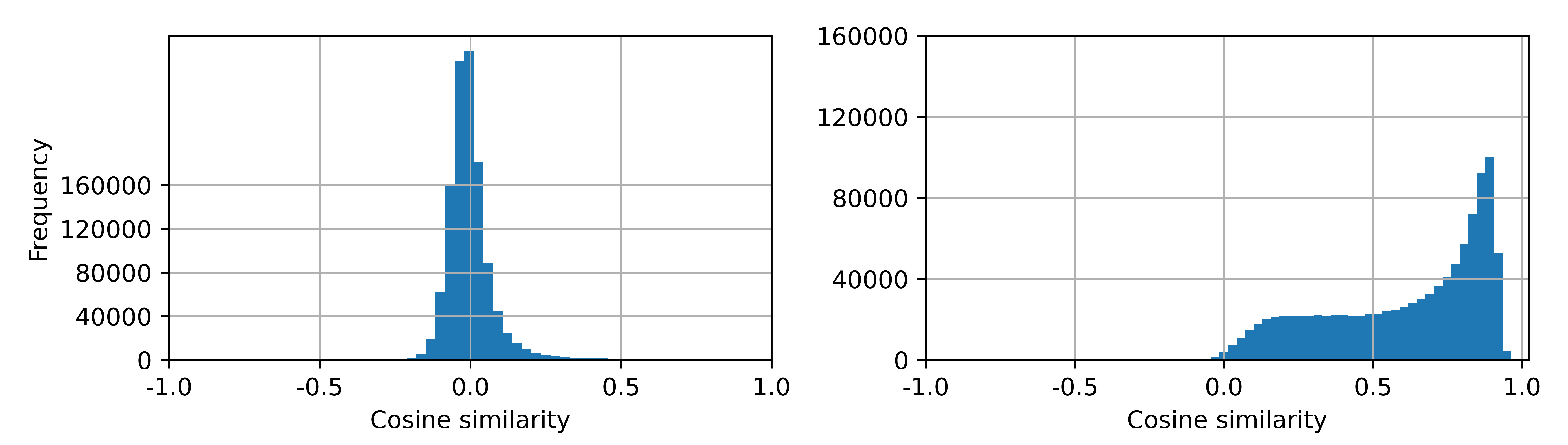}
		\caption{The distributions of cosine similarity after joint training: between the predicted and negative samples (left) and between the predicted and positive samples (right).}
		\vspace{-0.3cm}
		\label{fig:figure5}
	\end{figure}
	
 \textbf{Ablation study} In order to analyze the function of each module, we conduct ablation experiments and the corresponding results are shown in Table~\ref{tab:table2}.
	When only the regression loss function is utilized in the pre-training phase, the model is able to achieve a WER of 9.5 /15.7 on the validation/test set.
	In case the regression and contrastive loss functions (with $N = 100$ standard negative samples) are jointly optimized, the WER can be reduced, indicating that the joint training  can improve the performance of the pre-trained model in noisy scenarios.
	In case the regression and contrastive loss functions with non-semantic negative samples ($N$ = 50 standard negative samples and $N$ = 50 non-semantic negative samples) are jointly trained, the performance can be further improved, meaning that constructing non-semantic negative samples can improve the noise robustness.
	More importantly, it shows that the WER of the proposed model becomes smallest with an additional removal of more distinguishable negative samples ($k$ = 50). 
	The ASR model is then decoded jointly with the Transformer-based LM, resulting a WER of 3.1/5.8 on the validation/test set, which is the best performance in Table 1.
	
 \textbf{Visualization analysis} In addition, we analyze the impact of different numbers of standard negative samples and non-semantic negative samples on the WER, and the results are shown in Table~\ref{tab:table3}.
	It is clearly validated that introducing additional non-semantic negative samples in the pre-training phase can improve the ASR performance.
	However, it does not mean that the more non-semantic negative samples that are considered, the better the performance, as the choice of 50 standard negative samples and 50 non-semantic negative samples achieves the best performance.
	Finally, we visualize the distribution of positive and negative samples after the joint training of the models, which is shown in  Fig.~\ref{fig:figure5}.
	Compared with the bottom picture in Fig.~\ref{fig:figure3}, due to the increased difficulty of negative samples in the training process, the number of samples with zero cosine similarity decreases and the number of samples with unit cosine similarity increases, implying an increase in the discriminative capacity of the model, which can therefore obtain a more noise-robust representation.

	\vspace{-0.2cm}
	\section{CONCLUSION}
	\vspace{-0.15cm}
	\label{sec:conclusion}
	
	In this paper, we investigated the impact of constructing non-semantic negative samples in contrastive learning on the noise robustness of pre-trained data2vec.
	By dividing the features into patches at different sizes as negative samples, the noise robustness of the pre-training model can be improved.
	It was shown  that removing the more distinguishable negative samples enables a further ASR performance improvement.
	We found that jointly optimizing the contrastive learning and regression tasks can not only improve the performance, but also reduce the possibility of model collapse during the pre-training phase.

	
	\bibliographystyle{IEEEbib}
	\bibliography{strings,refs}
	
\end{document}